\documentclass[conference,anonymous=true]{IEEEtran}
\IEEEoverridecommandlockouts
\usepackage{cite}
\usepackage{amsmath,amssymb,amsfonts}
\usepackage{algorithmic}
\usepackage{graphicx}
\usepackage{textcomp}
\def\BibTeX{{\rm B\kern-.05em{\sc i\kern-.025em b}\kern-.08em
    T\kern-.1667em\lower.7ex\hbox{E}\kern-.125emX}}

\usepackage{subcaption}
\usepackage{multirow}
\usepackage{xspace}
\usepackage{url}
\usepackage{fancyhdr}
\usepackage{color}
\usepackage{hyperref}
\usepackage{balance}
\usepackage{booktabs}
\usepackage{enumitem}
\usepackage{amsthm, amssymb}
\theoremstyle{definition}

\newcommand{\modelname}{MVPRec\xspace}
\fancypagestyle{FirstPage}{
\lfoot{979-8-3503-2445-7/23/\$31.00 \copyright2023 IEEE}
}
\begin{document}

\title{Multi-view Graph Convolution for Participant Recommendation}


\author{\IEEEauthorblockN{Xiaolong Liu}
\IEEEauthorblockA{\textit{Department of Computer Science} \\
\textit{University of Illinois Chicago}\\
Chicago, USA \\
xliu262@uic.edu}
\and
\IEEEauthorblockN{Liangwei Yang}
\IEEEauthorblockA{\textit{Department of Computer Science} \\
\textit{University of Illinois Chicago}\\
Chicago, USA \\
lyang84@uic.edu}
\and
\IEEEauthorblockN{Chen Wang}
\IEEEauthorblockA{\textit{Department of Computer Science} 
\\
\textit{University of Illinois Chicago}\\
Chicago, USA \\
cwang266@uic.edu}
\and
\IEEEauthorblockN{Mingdai Yang}
\IEEEauthorblockA{\textit{Department of Computer Science} \\
\textit{University of Illinois Chicago}\\
Chicago, USA \\
myang72@uic.edu}
\and
\IEEEauthorblockN{Zhiwei Liu}
\IEEEauthorblockA{\textit{Salesforce AI Research} \\
Palo Alto, USA \\
zhiweiliu@salesforce.com}
\and
\IEEEauthorblockN{Philip S. Yu}
\IEEEauthorblockA{\textit{Department of Computer Science} \\
\textit{University of Illinois Chicago}\\
Chicago, USA \\
psyu@uic.edu}
}

\maketitle

\begin{abstract}

Social networks have become essential for people's lives. The proliferation of web services further expands social networks at an unprecedented scale, leading to immeasurable commercial value for online platforms. Recently, the group buying (GB) business mode is prevalent and also becoming more popular in E-commerce. GB explicitly forms groups of users with similar interests to secure better discounts from the merchants, often operating within social networks. It is a novel way to further unlock the commercial value by explicitly utilizing the online social network in E-commerce. Participant recommendation, a fundamental problem emerging together with GB, aims to find the participants for a launched group buying process with an initiator and a target item to increase the GB success rate. This paper proposes Multi-View Graph Convolution for Participant Recommendation (MVPRec) to tackle this problem. To differentiate the roles of users (Initiator/Participant) within the GB process, we explicitly reconstruct historical GB data into initiator-view and participant-view graphs. Together with the social graph, we obtain a multi-view user representation with graph encoders. Then MVPRec fuses the GB and social representation with an attention module to obtain the user representation and learns a matching score with the initiator's social friends via a multi-head attention mechanism. Social friends with the Top-$k$ matching score are recommended for the corresponding GB process. Experiments on three datasets justify the effectiveness of MVPRec in the emerging participant recommendation problem. MVPRec is open-sourced at \textcolor{blue}{\url{https://github.com/Xiaolong-Liu-bdsc/MVPRec}} to inspire further research in the new group buying E-commerce business mode.

\end{abstract}





\section{Introduction}
\thispagestyle{FirstPage}

The rapid developments of web services enlarge the expansion of social connections from various perspectives, such as information propagation~\cite{carminati2012trust} and work collaboration~\cite{wolf2008mining}.
Individuals, though
with far physical
distances, 
could establish connections through online social platforms.
As tools for everyday communication, social networks behind these platforms offer substantial commercial value~\cite{chen2014exploring} such as advertising revenue~\cite{tang2016profit} and data monetization~\cite{ngonmang2013monetization}. 
In E-commerce, social networks have already been used for user modeling to provide accurate recommendations~\cite{lin2019cross,yang2021consisrec, SCGRec, GraphRec, DiffNet, DiffNet++, CR-SoRec, socialgcn}. 

Recently, the landing of Temu~\footnote{\url{https://www.temu.com/}} in the US market raises research interests in the success of Pinduoduo~\footnote{\url{https://www.pinduoduo.com/}}, its origin counterpart from China. 
Pinduoduo distinguishes itself from other e-commerce entities through its innovative promotion of group buying (GB). This purchasing model, central to Pinduoduo's operations, consolidates consumers to purchase the same product, thereby leveraging merchant-offered discounts. This strategy not only resonates with the concept of economies of scale~\cite{stigler1958economies} but also presents a mutually beneficial scenario for both users and merchants. Merchants benefit from increased sales as the GB initiator effectively functions as a promoter, attracting additional buyers. Concurrently, consumers enjoy reduced prices.
In contrast to approaches that implicitly harness a user’s social network for user modeling~\cite{GraphRec,wu2020diffnet++}, GB explicitly capitalizes on shared interests among users to form groups, predominantly within social networks, aiming for greater discounts on specific items. This methodology represents a novel approach in harnessing the commercial potential of social networks for e-commerce platforms, significantly deviating from traditional models.
 
\begin{figure}[htbp]
    \centering
    \includegraphics[scale=0.40]{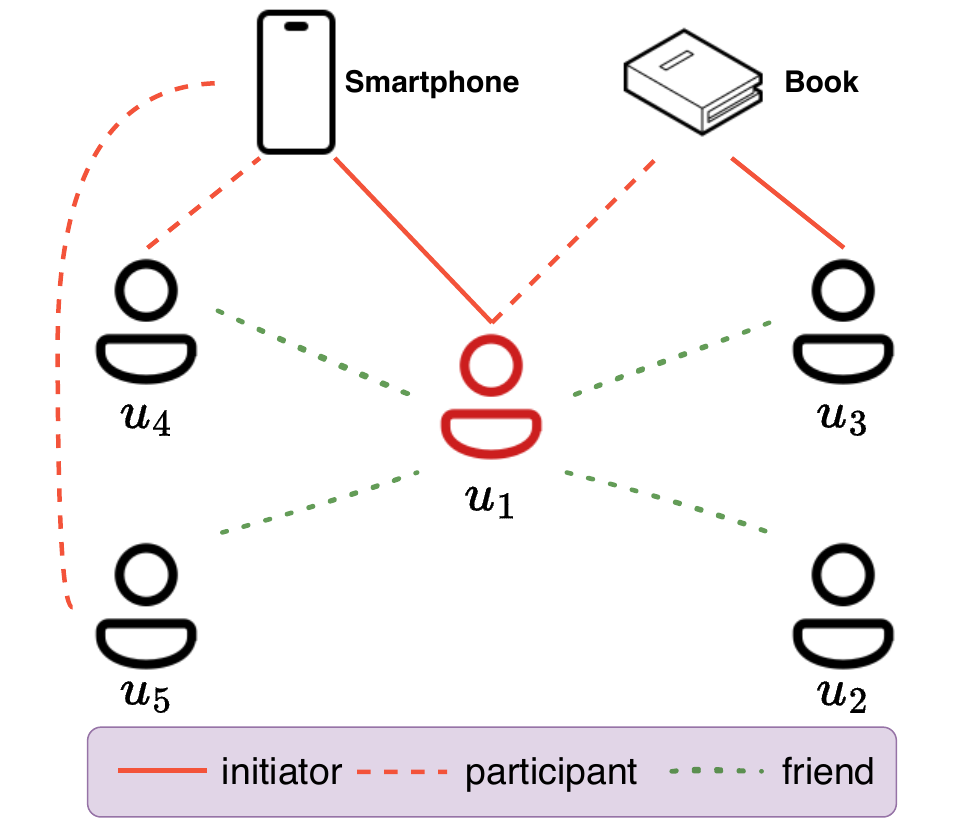}
    \caption{The illustration of the process of group buying. \textbf{Initiating Group Buying}: User $u_1$ initiates a group buying process and successfully forms a group with participants $u_4$ and $u_5$ to purchase a smartphone. \textbf{Participating in Group Buying}: User $u_1$ not only initiates a group buying process for a smartphone but also participates as a member in a group buying process initiated by $u_3$ to purchase a book.}
    \label{fig:mesh1}
\end{figure}


The process of GB is illustrated in Figure~\ref{fig:mesh1}. Each user can play two roles in a group buying process, i.e., initiator and participant. In this example, $u_1$ has four friends in the social platform (i.e., $u_2, u_3, u_4$, and $u_5$). As an initiator of the GB process, $u_1$ first finds a target item (smartphone) he/she would like to buy. Then, he/she launches the GB process and shares the item with friends to form groups. The GB will be successful if the group size exceeds the threshold the merchant set. Otherwise, he/she fails to get the item at a discounted rate. In this example, $u_1$ successfully forms groups with $u_4$ and $u_5$ to buy his desired item. In the same way, $u_1$ can also act as a participant of the GB process launched by $u_3$ to buy a book. In the GB process, the initiator plays the leading role in finding participants interested in his desired item while participants are not dominant roles. A natural question arises to complete a successful GB process: Which friend is willing to form groups with you to buy this item?

We term this question as a participant recommendation problem, which differs from previous problems utilizing social/group information. Social recommendation~\cite{GraphRec, SocialMF} enriches users' preferences by exploiting social homophily effect~\cite{mcpherson2001birds} (i.e., people are likely to build social relationships with those with similar preferences toward items). It aims to exploit user's social networks to recommend items to users rather than find the suitable participant. Besides, users play the same role within a social link but different roles in GB (Initiator/Participant).
Similar to social recommendation~\cite{consisrec,wu2020diffnet++, GraphRec, SocialMF}, group recommendation~\cite{AGREE, GroupIM, SGGCF, Consrec, ji2023multi, yin2020overcoming, jia2021hypergraph, liu2022federated}, also known as group decision-making, aims to recommend items to users based on users' enrolled groups of interest. It predicts the user's preference over items other than finding the participants of GB. At the same time, groups in group recommendation are usually long-term ones formed by users' common interests. However, groups in GB are usually short-term ones that vanish after the process, which can only take several minutes.

Building an effective participant recommendation model faces two challenges. 
(1) Role differentiation. In GB, a user's role is differentiated as either initiator or participant. The initiator is the leading role for the corresponding GB process. He/She finds the interested item, launches the GB process, and finds suitable participants. In comparison, participants only need to consider whether to join the group. Different users play different roles in a single GB process, and one initiator user may become the participant in another GB launch. Properly differentiating the roles during modeling is the first distinctive challenge in the participant recommendation problem.
(2) Heterogeneous information fusion. A GB process involves data from both social networks and historical GB cases. Social networks are formed between users to indicate their relationships. One historical GB case involves at least two users and one item. Social networks and GB processes exhibit different information, and how to tackle the information heterogeneity is the second challenge.

In this paper, we propose \textbf{M}ulti-\textbf{V}iew Graph Convolution for \textbf{P}articipant \textbf{Rec}ommendation (\modelname) to find suitable participants among user's social friends for an intended group-buying process. To differentiate the user roles during the GB process, we explicitly reconstruct historical GB data into an initiator-view graph and a participant-view graph. After the Light Graph Convolution~\cite{LghtGCN}, \modelname obtains both initiator-view and participant-view dense user representation. To fuse the heterogeneous information, \modelname further encodes the social graph to obtain the user's social representation and then fuses it with initiator/participant-view representation via an attention module. \modelname recommends participants based on both the initiator and the target item information. We propose fusing initiator and target item information into a query vector to match the suitable participants within the user's social friends. Then the Top-$k$ participants are recommended with the highest $k$ matching score calculated by a participant prediction module. Our main contributions are summarized as follows:
\begin{itemize}
    \item Conceptually, we are among the earliest attempts to study the participant recommendation problem, a new task brought by the group buying E-commerce mode.
    \item Methodologically, we identify the challenges behind the problem and propose \modelname that can effectively find suitable participants for a launched group buying process with initiator and target item.
    \item Experimentally, we conduct extensive experiments on three datasets to validate the effectiveness of \modelname on participant recommendation task.
\end{itemize}

\section{Data Analysis}
\begin{figure*}
     \centering
     \captionsetup{justification=centering}
     \begin{subfigure}[b]{0.325\textwidth}
         \centering
         \includegraphics[width=\textwidth]{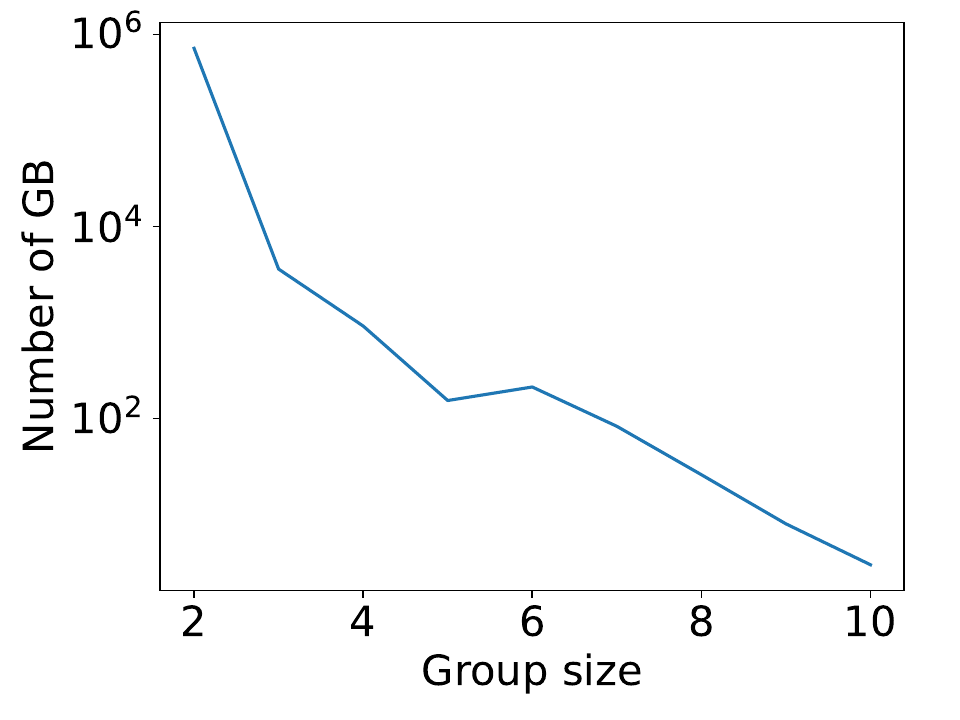}
         \caption{Number of group buying with regard to group size.}
         \label{group size}
     \end{subfigure}
     \hfill
        \begin{subfigure}[b]{0.325\textwidth}
         \centering
         \includegraphics[width=\textwidth]{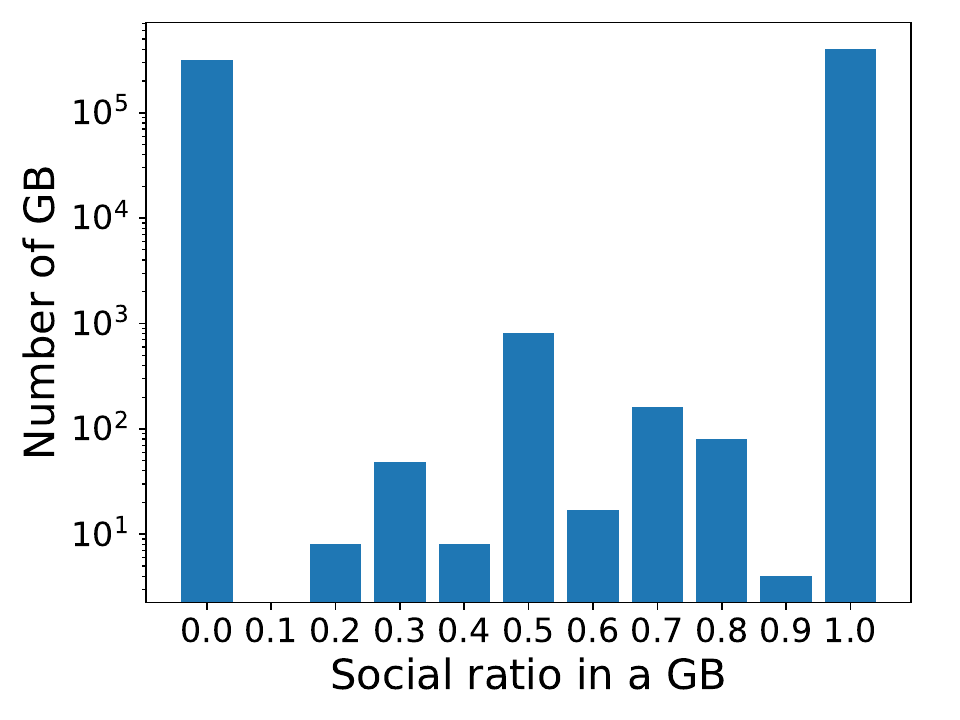}
         \caption{Distribution of group buying within the social network.}
         \label{social role}
     \end{subfigure}
      \hfill
    \begin{subfigure}[b]{0.325\textwidth}
         \centering
         \includegraphics[width=\textwidth]{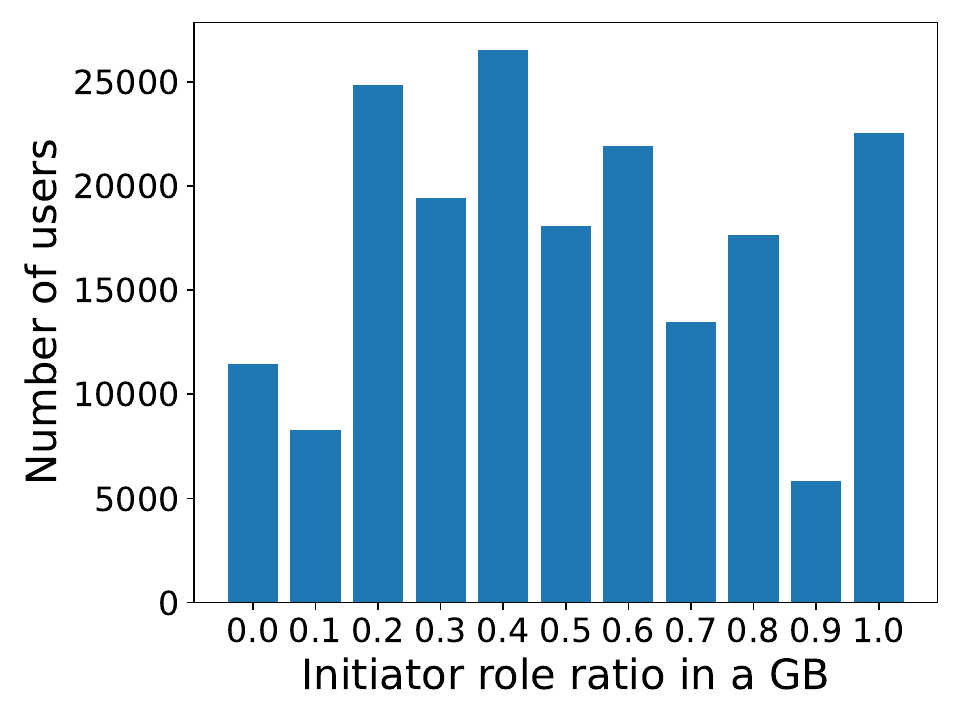}
         \caption{Distribution of user initiator-participant role ratio.}
         \label{role ratio}
     \end{subfigure}
      \hfill
\end{figure*}

In this section, we present an overview of the GB data statistics.
BeiBei~\cite{GBGCN}, an online retail platform based in China, serves as the only presently accessible dataset for scholarly investigation in the domain of group buying.

\subsection{Number of groups buying with regard to group size.}

In our initial analysis, we examined the distribution of group sizes in each GB instance, as illustrated in Figure~\ref{group size}. It was observed that a significant majority of GB cases are characterized by groups comprising merely two individuals, suggesting a typical scenario involving one initiator and one participant. Furthermore, there is a noticeable inverse correlation between the size of the group and the frequency of GB occurrences, indicating a prevalent trend of smaller group sizes within this context. This tendency towards smaller groups underscores the relevance of prioritizing the top few recommended participants in evaluating the effectiveness of the model in practical applications. Consequently, this informed our decision to employ the Top-3 ranking metric for the evaluative purposes in the experimental section detailed in Sec.~\ref{sec:rq1}.

\subsection{Distribution of group buying within the social network.}

In Figure~\ref{social role}, the x-axis represents the initiator's social friend proportion in relation to the total participant count in a GB scenario. The social ratio, quantified as 1, denotes scenarios where all participants are within the initiator's social circle, while a ratio of 0 indicates a complete lack of social connections among the participants. Empirical analysis reveals that in most GB interactions, especially in cases where the social ratio is greater than 0.5, the participants tend to be within the initiator's network of social acquaintances. Given this pattern, our approach primarily focuses on recommending participants from within the initiator's social network in order to effectively manage this phenomenon at a larger scale.

\subsection{Distribution of user initiator-participant role ratio.}

In the context of GB interactions, users engage in distinct roles, alternating between being initiators and participants. This dynamic is quantitatively represented in Figure~\ref{role ratio}, which illustrates the user initiator-participant role ratio. A ratio of 1 in this context is indicative of a user predominantly assuming the initiator role, whereas a ratio of 0 correlates with the user consistently participating rather than initiating. The distribution pattern suggests that users typically undertake a mix of both roles. This role variation is crucial, as it facilitates the differentiation of user roles in GB, thereby enabling a more nuanced understanding of their respective preferences.

Building upon this comprehensive data analysis, we have developed an innovative model tailored to enhance the accuracy of participant recommendations. This model is particularly attuned to the specific characteristics of the data under consideration.

\section{Preliminary}\label{sec:pre}
In this section, we illustrate the preliminaries to understand \modelname, which includes task formulation and light graph convolution.

\subsection{Task Formulation}
We formally formulate the problem of participant recommendation task with group buying interactions.
The problem is defined on a set of users $\mathcal{U} = \{u_1,u_2,...,u_{\left | \mathcal{U} \right|}\}$ and a set of items $\mathcal{I} = \{i_1,i_2,...,i_{\left | \mathcal{I} \right|}\}$.
A social network is denoted as $\mathcal{S} \in \mathbb{R}^{|\mathcal{U}| \times |\mathcal{U}|}$, where $\mathcal{S}_{mn} = 1$ represents there is a social connection between $u_m$ and $u_n$, and we utilize $\mathcal{S}_u$ to denote the set of social friends of user $u$.
We further define the group buying interactions by $\mathcal{P} = \{(u, i, \mathcal{U}_p)| u \in \mathcal{U}, i \in \mathcal{I}, \mathcal{U}_p \subseteq \mathcal{U} \}$, where the initiator $u$ purchases the item $i$ with a set of $k$ participants $\mathcal{U}_p = \{ u_{p1}, ..., u_{pk} \}$.
Given an initiator $u$ and a target item $i$, the participant recommendation aims to retrieve a social friend of $u$ who is interested in $i$ to form the group to buy the target item.

\subsection{Light Graph Convolution (LGCN)}\label{sec:lgcn}
\modelname models both the social network and historical group buying data as graphs. We use the Light Graph Convolution (LGCN)~\cite{LghtGCN} as the encoder to obtain user/item representation from different subgraphs. It is widely used due to its efficiency. LGCN($\ast$) convolutes over graphs with the message-passing mechanism to aggregate the neighborhood's information to the center node, which is defined as:
\begin{equation}
    \begin{gathered}
        \mathbf{e}_u^{(k+1)} = \sum_{v \in \mathcal{N}_u} \frac{1}{\sqrt{\left | \mathcal{N}_u \right |} \sqrt{\left | \mathcal{N}_i \right |} } \mathbf{e}_{i}^{(k)}, \\
        \mathbf{e}_{i}^{(k+1)} = \sum_{u \in \mathcal{N}_i} \frac{1}{\sqrt{\left | \mathcal{N}_i \right |} \sqrt{\left | \mathcal{N}_u \right |} } \mathbf{e}_u^{(k)},
    \end{gathered}
\end{equation}
where $\mathbf{e}_u^{(k)}, \mathbf{e}_{i}^{(k)}$ denote the latent vectors of user $u$ and item $i$ after $k$ layers propagation. $\mathcal{N}_u$ denotes the set of items that interact with user $u$, and $\mathcal{N}_i$ denotes the set of users interacting with item $i$.

Following the completion of $K$ layers of propagation, these embeddings acquired at each layer are combined to create the ultimate embedding for each user and item:
\begin{align}
    \begin{aligned}
        \mathbf{e}_u = \sum_{k=0}^{K} \mathbf{e}_u^{(k)},
        \mathbf{e}_i = \sum_{k=0}^{K} \mathbf{e}_i^{(k)}.
    \end{aligned}
\end{align}
$\mathbf{e}_u$ and $\mathbf{e}_i$ are the encoded user/item representation with the LGCN($\ast$).
\begin{figure*}[htbp]
    \centering
    \includegraphics[scale=0.33]{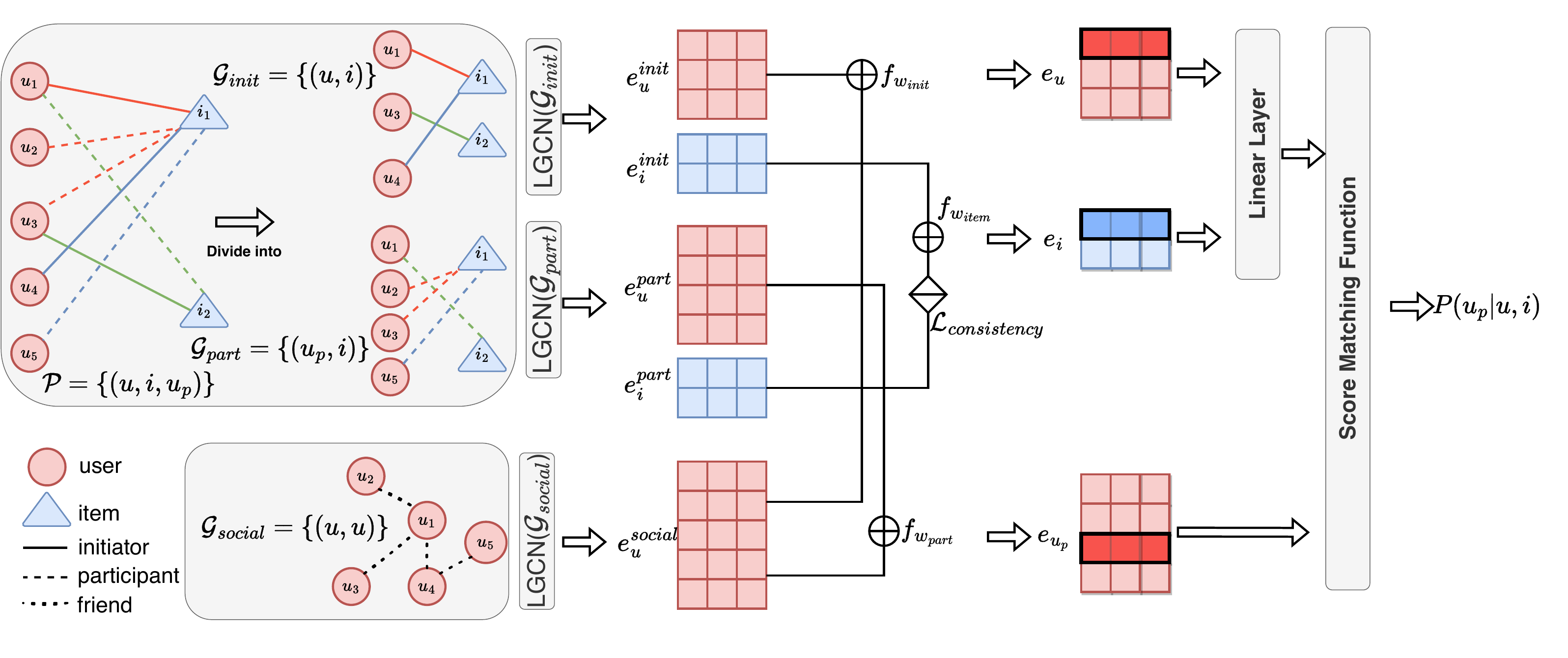}
    \caption{Our multi-views propagation framework comprises three key components: \textbf{1.Multi-View Partitioning:} We partition the group buying graph $\mathcal{P}$ into initiator view $\mathcal{G}{init}$ and participant view $\mathcal{G}{part}$ to simplify complex relationships. Color-coded events represent group buying instances, and we obtain user and item embeddings through LGCN aggregation. \textbf{2.Incorporating Social Networks:} Social network data from $\mathcal{G}_{social}$ enhances user embeddings in both initiator and participant views, creating ultimate user representations. We also handle variations in item embeddings between these views, fusing them using a fusion function and consistency loss. \textbf{3.Scoring and Prediction:} Utilizing ultimate user embeddings and a unified item embedding table, we apply a participant prediction module to predict user $u_p$'s preferences for a given initiator user $u$ and target item $i$.}
    \label{model}
\end{figure*}

\section{Proposed Model: \modelname}
In this section, we illustrate the novel framework \modelname in detail. The framework is shown in Fig.~\ref{model}. It consists of the Embedding Layer, Multi-view Learning, Participant Prediction, and consistency loss regularization.

\subsection{Embedding Layer}
In the same way words and phrases are represented through deep learning techniques, embeddings are also commonly employed in recommender systems, as demonstrated in works such as LightGCN~\cite{LghtGCN} and MF-BPR~\cite{rendle2012bpr}. An embedding layer functions as a lookup table that associates user or item IDs with compact, dense vectors:
\begin{equation}
    \textbf{E}^{(0)}=\left(\textbf{e}_1^{(0)}, \textbf{e}_2^{(0)}, \ldots, \textbf{e}_{|\mathcal{U}|+|\mathcal{I}|}^{(0)}\right), \\
\end{equation}
where $\mathbf{e}^{(0)}\in \mathbb{R}^d$ is the $d$-dimensional dense vector for user/item. Subsequently, an embedding retrieved from the embedding table is introduced into a Graph Neural Network (GNN) for the purpose of information aggregation. Thus, the embedding retrieved from the embedding table is represented as $\mathbf{e}_i^{(0)}$ to denote the representation before the first layer of the model.

\subsection{Multi-view learning}
A user can act as an initiator or a participant in the group buying process.
Different roles play different parts in the process. An initiator finds the target item and calls on his/her friends to form groups to satisfy the group discount requirement from the merchants while a participant only needs to consider whether to accept the proposal from the initiator.

To reflect the role difference, \modelname reconstructs the group buying interactions $\mathcal{P}=\{(u,i,\mathcal{U}_p)|u\in\mathcal{U},i \in \mathcal{I}, \mathcal{U}_p \subseteq \mathcal{U} \}$ into two subgraphs $\mathcal{G}_{init} = \{ (u, i)| u \in \mathcal{U}, i \in \mathcal{I} \}$ and $\mathcal{G}_{part} = \{ (u_p, i)| u_p \in \mathcal{U}_p,  i \in \mathcal{I}\}$. $\mathcal{G}_{init}$ is the initiator-item bipartite graph that models the initiator-view information, and $\mathcal{G}_{part}$ is the participant-item bipartite graph to model the participant-view information. Besides, we build the social graph $\mathcal{G}_{social} \in \mathbb{R^{\left| \mathcal{U} \right| \times \left| \mathcal{U} \right|}}$ from the social network $\mathcal{S}$ to generate the social-view user representation. Then different views of user/item embedding are obtained by encoding these three graphs with LGCN~\cite{LghtGCN} introduced in Sec.~\ref{sec:lgcn} as:
\begin{align}
    \begin{aligned}
          \mathbf{e}_u^{init}, \mathbf{e}_i^{init} &= \text{LGCN} (\mathcal{G}_{init}),\\
          \mathbf{e}_{u_p}^{part}, \mathbf{e}_i^{part} &= \text{LGCN} (\mathcal{G}_{part}),\\
          \mathbf{e}_u^{social} &= \text{LGCN}(\mathcal{G}_{social}),
    \end{aligned}
\end{align}
where $\mathbf{e}_u^{init}$ is the initiator-view user representation, and $\mathbf{e}_i^{init}$, $\mathbf{e}_{u_p}^{part}$, $\mathbf{e}_i^{part}$, $\mathbf{e}_u^{social}$ are illustrated in the same way.
For $\mathcal{G}_{social}$, LGCN performs convolution between social edges similarly to the user-item bipartite graph. Separating the learning of different information directly into different views makes \modelname more easily to capture the differences of user roles in the GB process.

We then design a fusion function $f_w(\mathbf{e}_1, \mathbf{e}_2)$ to adaptively fuse information from different views as:
\begin{align}
    \begin{aligned}
    \textbf{e}_{fuse} &= f_\mathbf{w} (\mathbf{e}_1, \mathbf{e}_2)\\
    &= \frac{\mathbf{w} \cdot \mathbf{e}_1}{\mathbf{w} \cdot \mathbf{e}_1 + \mathbf{w} \cdot \mathbf{e}_2} \cdot \mathbf{e}_1 + \frac{\mathbf{w} \cdot \mathbf{e}_2}{\mathbf{w} \cdot \mathbf{e}_1 + \mathbf{w} \cdot \mathbf{e}_2} \cdot \mathbf{e}_2,
    \end{aligned}
\end{align}
where $\cdot$ is the dot product, and $\mathbf{w} \in \mathbb{R}^{d}$ is the parameter necessary in both numerator and denominator to adaptively fuse information from different views.
Then the initiator, participant and item representation are obtained by different fusion functions:
\begin{align}
    \begin{aligned}
      \textbf{e}_{u} &= f_{\textbf{w}_{init}} (\textbf{e}_u^{init}, \mathbf{e}_u^{social}), \\
      \textbf{e}_{u_p} &= f_{\textbf{w}_{part}} (\textbf{e}_{u_p}^{part}, \mathbf{e}_{u_p}^{social}), \\
      \textbf{e}_i &= f_{\textbf{w}_{item}} (\textbf{e}_i^{init}, \mathbf{e}_i^{part}).
    \end{aligned}
\end{align}
In \modelname, we fine-grain user representation as initiator-view $\textbf{e}_{u}$ and participant-view $\textbf{e}_{u_p}$ to differentiate the user roles in the GB process. In one GB process, we use $\textbf{e}_{u}$ if $u$ acts as the initiator, and $\textbf{e}_{u_p}$ if $u$ is the participant.

\subsection{Participant Prediction}
After we obtain different views of user representation, \modelname predicts the participant of the GB process with an initiator and target item. We assume the participant is conditioned on both the initiator and target item and design a linear layer to obtain the query embedding by fusing both the information:
\begin{align}
    \mathbf{q} = \sigma(\mathbf{W} [\mathbf{e}_{u} || \mathbf{e}_i]),
    \label{query}
\end{align}
where $||$ is the concatenation function, $\mathbf{W} \in \mathbb{R}^{d*2d}$ is a linear transformation to keep query embedding $\mathbf{q} \in \mathbb{R}^{d}$, and $\sigma$ is the sigmoid function. Then we compute the likelihood of each participant on the query in a multi-head manner~\cite{attention} as: 
\begin{align}
     P(u_p|u, i) = \frac{1}{H}\sum_{h=1}^{H}\frac{\text{exp}(\textbf{W}_q^{h} \textbf{q} \cdot \textbf{W}_k^{h} \textbf{e}_{u_p})}{\text{exp}(\sum_{\forall u_{p_j} \in \mathcal{S}_u} \textbf{W}_q^{h} \textbf{q} \cdot \textbf{W}_k^{h} \textbf{e}_{u_{p_j}})},
\end{align}
where $\textbf{W}_q^{h} \in \mathbb{R}^{d*d}$ and $\textbf{W}_k^{h} \in \mathbb{R}^{d*d}$ is the query/participant transformation parameter on the $h$-th head.
Multi-head attention mechanism allows the model to pay attention to different parts of the input sequence simultaneously.
Each head can focus on different aspects of the context. Therefore, the intricate collaborative signals of potential participants could be captured in this way.
Then the participant recommendation loss function $\mathcal{L}_{part}$ is defined as the negative log likelihood:
\begin{align}
    \mathcal{L}_{part} = - \sum_{(u, i, \mathcal{U}_p) \in \mathcal{P}} \sum_{\forall u_p \in \mathcal{U}_p} \text{ln} P(u_p|u, i).
\end{align}
By minimizing $\mathcal{L}_{part}$, we increase the likelihood of the ground-truth participant given the initiator and target item.

\subsection{Consistency loss}
$\mathbf{e}_i^{init}$ and $\mathbf{e}_i^{part}$ are different views for item obtained from $\mathcal{G}_{init}$ and $\mathcal{G}_{part}$. In a GB process, both the initiator and participant show their interest in the target item. We assume item representation should reveal their common interest, and $\mathbf{e}_i^{init}$ and $\mathbf{e}_i^{part}$ should not be differentiated in a large margin.
Thus, we further add a consistency loss to regularize $\mathbf{e}_i^{init}$ and $\mathbf{e}_i^{part}$ to be similar to each other:
\begin{align}
    \begin{aligned}
       \mathcal{L}_{consistency} = -\frac{\mathbf{e}_{i}^{init} \cdot \mathbf{e}_{i}^{part}}{||\mathbf{e}_i^{init}||_{2} \cdot ||\mathbf{e}_i^{part}||_{2}}.
    \end{aligned}
\end{align}
Then the final loss for optimization is obtained as:
\begin{align}
    \begin{aligned}
       \mathcal{L} = \mathcal{L}_{part} + \lambda_1\mathcal{L}_{consistency} +\lambda_2 {\left \| \Theta \right \|}_2^2,
    \end{aligned}
\end{align}
where $\Theta$ represents all parameters within \modelname. $\lambda_1$ and $\lambda_2$ are hyper-parameters to balance the loss weights.

\section{Experiments}
In this section, we conduct extensive experiments on three datasets to answer the following research questions (RQs):
\begin{itemize}
    \item \textbf{RQ1}: Does \modelname outperform existing methods in the participant recommendation problem?
    \item \textbf{RQ2}: Are different components in \modelname necessary to increase the performance on the participant recommendation problem?
    \item \textbf{RQ3}: How do the hyper-parameters influence \modelname?
\end{itemize}

\subsection{Experiment Settings}
\subsubsection{Dataset}
\begin{table}
  \caption{Statistics of the Datasets}
  \label{dataset}
  \centering
  \scalebox{1.2}{
  \begin{tabular}{l c c c}
        \toprule
        \textbf{Dataset} & \textbf{BeiBei} & \textbf{Ciao} & \textbf{Epinion} \\
        \hline
        \textbf{\#Users} & 186,352 & 2,342 & 18,089\\

        \textbf{\#Items} & 30,543 & 77,540 & 261,649\\

        \textbf{\#Social} & 675,911 & 57,544 & 355,813\\

        \textbf{\#GB} & 403,419 & 15,138 & 69,284\\

        \bottomrule
  \end{tabular}
  }
\end{table}
In this section, we introduce our experiments on the public dataset: BeiBei, Ciao, and Epinion. 
The group buying dataset BeiBei, a Chinese infant product platform, is released by~\cite{GBGCN}. 
We randomly split the group buying interactions of each initiator into the training set (80\%), the validation set (10)\%), and the test set (10\%).
Owing to the scarcity of datasets specifically tailored for group buying analysis, we developed two additional datasets from Ciao and Epinions, which are well-established in the domain of social recommendation research. In these datasets, a purchase made by a user and their friends involving a common item is classified as group buying activity. The user who first purchases the item is designated as the initiator. Subsequently, friends who acquire the same item following the initiator are identified as participants in this group buying activity.
The detail can be seen in Table~\ref{dataset}.

\begin{table*}[t]
    \caption{Comparison of Top-K performance on three datasets with baselines in participant recommendation problem. The best and second-best results are in bold and underlined, respectively}\label{tab:performance}.
    \label{result}
    \small
    \centering
    \setlength{\tabcolsep}{3mm}{
    \scalebox{1.0}{
    \begin{tabular}{l|l|cccccccc}
         \hline
         Dataset & Metric & User-activity & Cosine Similarity & MF & TransE & RGCN & J2PRec & \modelname & Improv. \\
         \hline
         \multirow{5}{*}{BeiBei}&N@1& 
         0.0946 &
        0.0504 &
          0.1336 &
          \underline{0.1390} &
          0.1342 &
          0.1209&
           \textbf{0.1396} & 0.43\%
          \\
         & N@2 & 
         0.1545 &
         0.0722 &
         0.2145 &
         \underline{0.2188} &
         0.2106 &
         0.1974 &
         \textbf{0.2252} & 2.93\%
         \\

        & N@3 & 
         0.2054 &
         0.1101 &
         0.2779 &
         \underline{0.2810} &
         0.2684 &
         0.2585 &
          \textbf{0.2915} & 3.74\% \\
         
         & R@2 & 
         0.1896 &
         0.0850 &
         0.2619 &
         \underline{0.2655} &
         0.2553 &
         0.2422 &
          \textbf{0.2753} & 3.69\% \\

         & R@3 & 
         0.2914 &
         0.1607 &
         0.3884 &
         \underline{0.3899} &
         0.3709 &
         0.3644 &
          \textbf{0.4078} & 4.59\% \\

         \hline
         \multirow{5}{*}{Ciao}&N@1& 
         0.1079 &
        0.1186 &
          \underline{0.1532} &
          0.1506 &
          0.0726 &
          0.1266 &
           \textbf{0.1972} & 28.72\%
          \\

        & N@2 & 
         0.1609 &
         0.1779 &
         \underline{0.2377} &
         0.2044 &
        0.1067 &
         0.1926 &
         \textbf{0.2695} & 13.38\%
         \\

     & N@3 & 
         0.1975 &
         0.2222 &
         \underline{0.2873} &
         0.2373 &
         0.1306 &
         0.2335 &
         \textbf{0.3121} & 8.63\% \\
         
         & R@2 & 
         0.1919 &
         0.2125 &
         \underline{0.2871} &
         0.2358 &
         0.1266 &
        0.2312 &
        \textbf{0.3118} & 8.60\%
         \\

         & R@3 & 
         0.2652 &
         0.3011 &
         \underline{0.3863} &
         0.3018 &
         0.1746 &
         0.3131 &
         \textbf{0.3971} & 2.80\%
          \\
         \hline
         \multirow{5}{*}{Epinion}&N@1& 
         0.1674 &
        \underline{0.2575} &
          0.2398 &
          0.1603 &
          0.0700 &
          0.1601&
           \textbf{0.3225} & 34.49\%
          \\
          
                  & N@2 & 
         0.2369 &
         0.2945 &
         \underline{0.3028} &
         0.1929 &
         0.1074 &
         0.2174 &
         \textbf{0.3755} & 24.01\%
         \\

                 & N@3 & 
         0.2651 &
         0.3168 &
         \underline{0.3366} &
         0.2106 &
         0.1242 &
         0.2331 &
         \textbf{0.4056} & 20.50\%
          \\
         
         & R@2 & 
         0.2775 &
        0.3162 &
         \underline{0.3396} &
         0.2120 &
        0.1292 &
         0.2508 &
         \textbf{0.4065} & 19.70\%
          \\
         
         & R@3 & 
         0.3340 &
         0.3609 &
         \underline{0.4072} &
         0.2474 &
         0.1628 &
         0.2824 &
         \textbf{0.4667}  & 14.61\%
          \\

         \hline
    \end{tabular}
    }}
\end{table*}

\subsubsection{Baselines}
To demonstrate the effectiveness of our model, we compare \modelname with several baselines on three datasets. 
To our knowledge, J2PRec\cite{J2PRec} stands as the sole baseline that closely aligns with participant recommendations. 
As several social and group recommender systems, such as those presented in J2PRec, are primarily geared toward offering personalized recommendations to users or groups. Moreover, the formation of most group-buying records is occasionally, they don't directly align with our current task. Consequently, we've made minor modifications to other baseline methods to better adapt them to our specific requirements.

\begin{itemize}[leftmargin=*]
    \item User-activity: ranks the social friends according to their activities, measured by the number of purchased records.
    \item Cosine Similarity: retrieves the most similar friends with the initiator.
    \item MF~\cite{MF}: employs matrix factorization to decompose the user-item interaction matrix into latent vectors for users and items by utilizing the Bayesian Personalized Ranking (BPR) loss function. 
    \item TransE~\cite{TransR}: a knowledge graph embedding model designed to represent entities and relationships as continuous vectors, enabling the modeling of semantic relationships in structured data. 
    \item RGCN~\cite{RGCN}: a deep learning model tailored for processing and reasoning over knowledge graphs by incorporating graph convolutional operations.
    \item J2PRec~\cite{J2PRec}: is a joint product-participant recommendation model that learns user and item embedding by iteratively aggregating from the relational group-buying graph.
\end{itemize}

To adapt MF to the participant recommendation task, we introduce a dot product operation between the item vector and participant vector to determine whether to recommend a particular participant for a specific group buying, without taking the initiator into account.
For the utilization of TransE and RGCN, we introduce five relations (buy\_as\_initiator, buy\_as\_participant, bought\_by\_initiator, bought\_by\_participant, friend\_of) along with their corresponding interactions to construct the relational graph. The final outcome is evaluated using a score function applied to $(u_p, \text{buy\_as\_participant}, i)$.

\subsubsection{Evaluation Method}
In the evaluation stage, we rank the ground-truth participants with all friends of the initiator.
For effectiveness, we adopt Recall@$k$ and NDCG@$k$ as the metrics by setting $k \in \{1, 2, 3\}$, which aligns with real-world applications where the top matches are of utmost importance.

\subsubsection{Hyper-parameter Setting}
For all baselines and our model, we fix the embedding size as 32 with the Adam optimizer. 
Grid search is conducted to tune the hyperparameters in \modelname.
The learning rate is searched in \{1e-1, 5e-2, 1e-2, 5e-3, 1e-3, 5e-4, 1e-4\} and weight decay is tuned in \{1e-3, 1e-4, 1e-5, 1e-6\} for all models.
Since the number of LGCN layers $K$, the coefficient on consistency loss $\lambda_1$, and the number of heads $L$ are the unique parameters in our model, we tune them in \{1, 2, 3, 4, 5\}, \{0.2, 0.4, 0.6, 0.8, 1.0\}, and \{1, 2, 3, 4, 5\}, respectively.

\begin{figure*}[t]
     \centering
     \begin{subfigure}[b]{0.24\textwidth}
         \centering
         \includegraphics[width=\textwidth]{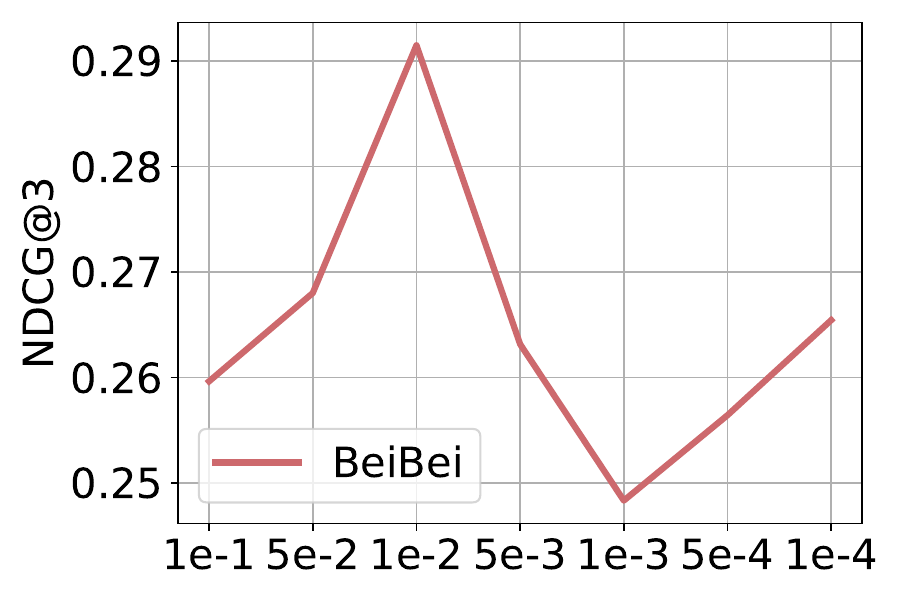}
         \caption{Learning Rate $lr$}
         \label{BeiBei lr}
     \end{subfigure}
     \hfill
        \begin{subfigure}[b]{0.24\textwidth}
         \centering
         \includegraphics[width=\textwidth]{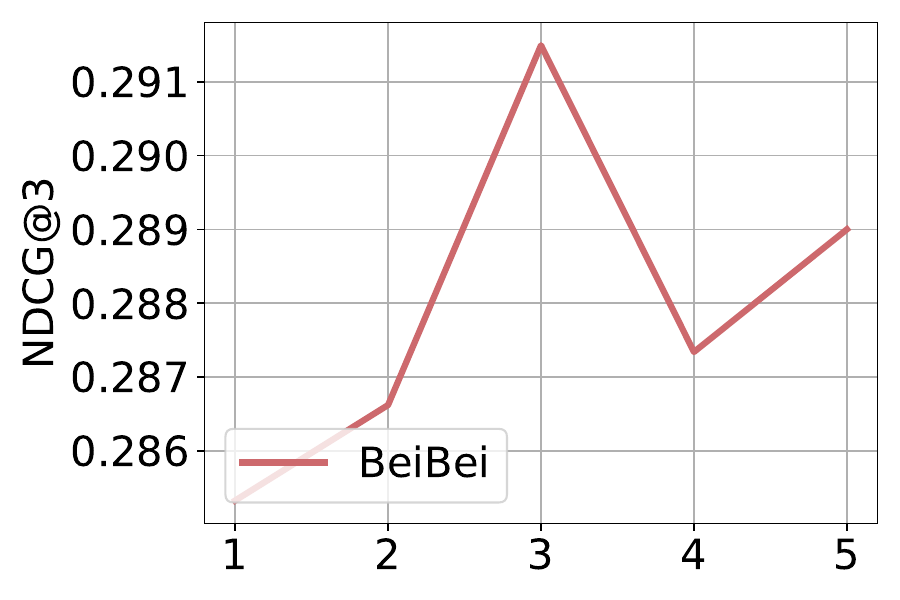}
         \caption{Number of layers $K$}
         \label{BeiBei layer}
     \end{subfigure}
      \hfill
    \begin{subfigure}[b]{0.24\textwidth}
         \centering
         \includegraphics[width=\textwidth]{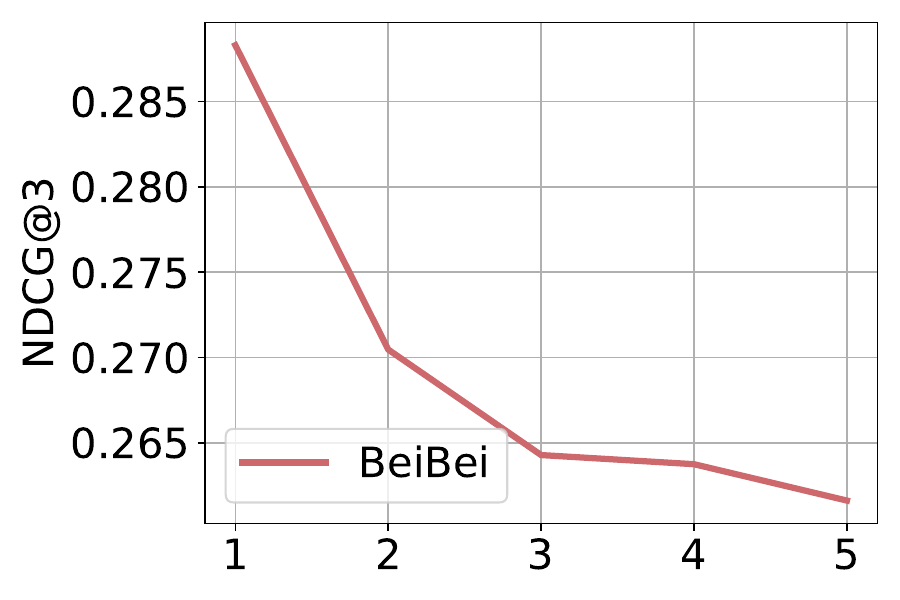}
         \caption{Number of heads $H$}
         \label{BeiBei heads}
     \end{subfigure}
      \hfill
    \begin{subfigure}[b]{0.24\textwidth}
         \centering
         \includegraphics[width=\textwidth]{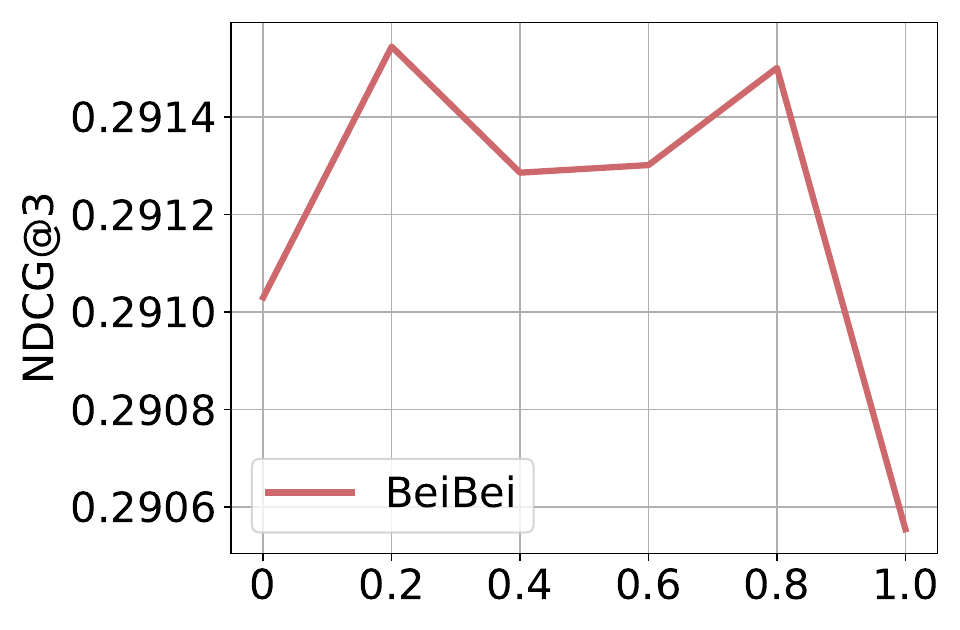}
         \caption{Regularization $\lambda_1$}
         \label{BeiBei reg}
     \end{subfigure}

    \begin{subfigure}[b]{0.24\textwidth}
         \centering
         \includegraphics[width=\textwidth]{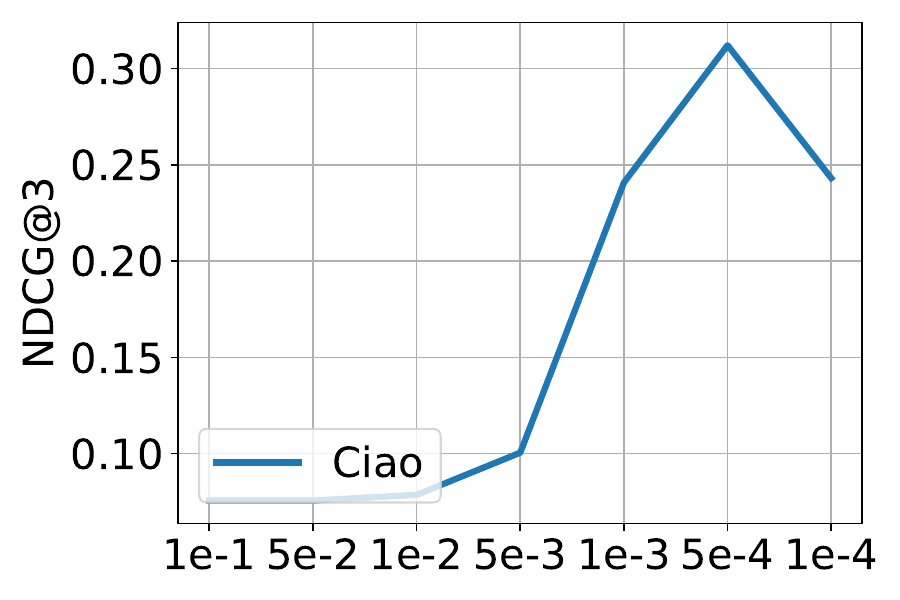}
         \caption{Learning Rate $lr$}
         \label{Ciao lr}
     \end{subfigure}
     \hfill
     \begin{subfigure}[b]{0.24\textwidth}
         \centering
         \includegraphics[width=\textwidth]{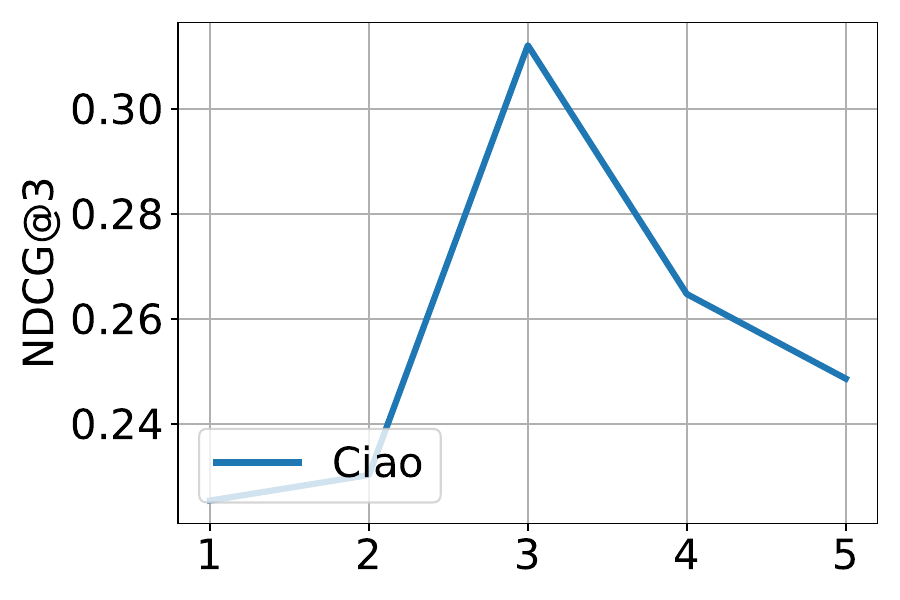}
         \caption{Number of layers $K$}
         \label{Ciao layer}
     \end{subfigure}
     \hfill
     \begin{subfigure}[b]{0.24\textwidth}
         \centering
         \includegraphics[width=\textwidth]{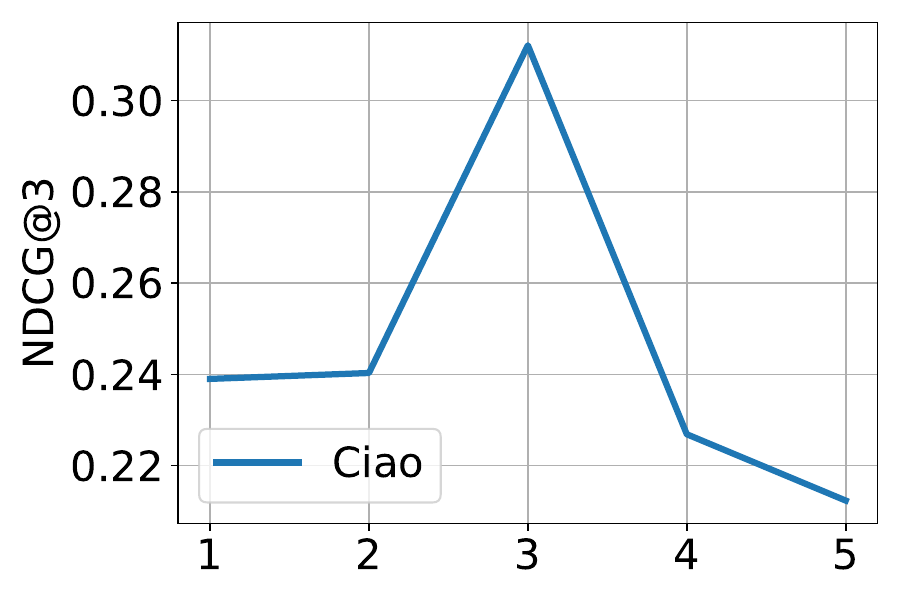}
         \caption{Number of heads $H$}
         \label{Ciao heads}
     \end{subfigure}
          \hfill
     \begin{subfigure}[b]{0.24\textwidth}
         \centering
         \includegraphics[width=\textwidth]{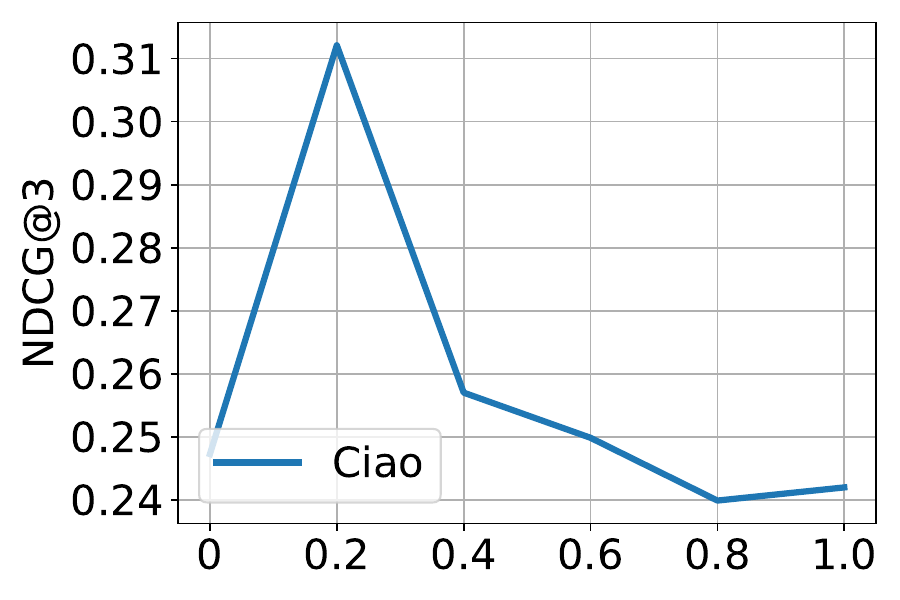}
         \caption{Regularization $\lambda_1$}
         \label{Ciao reg}
     \end{subfigure}

    \begin{subfigure}[b]{0.24\textwidth}
         \centering
         \includegraphics[width=\textwidth]{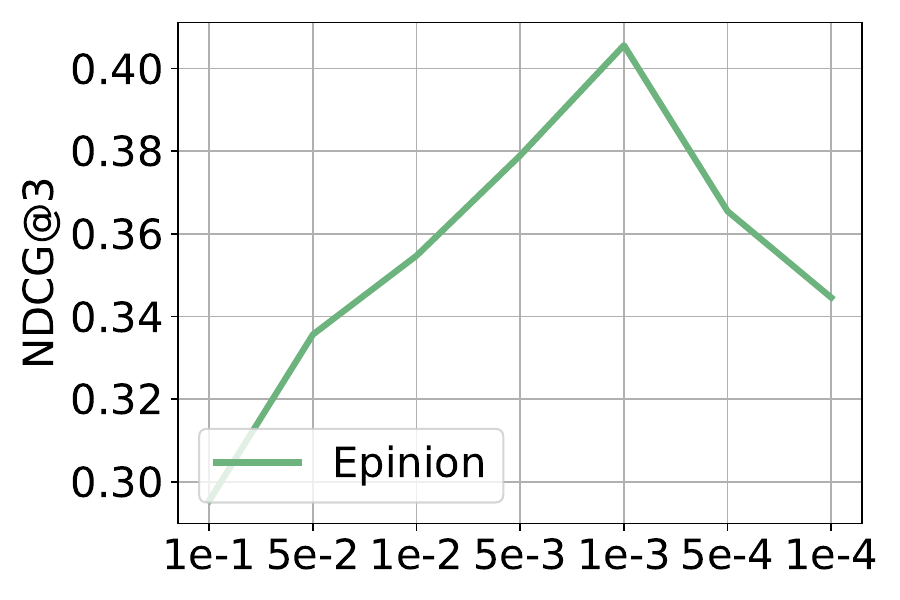}
         \caption{Learning Rate $lr$}
         \label{Epinion lr}
     \end{subfigure}
     \hfill
     \begin{subfigure}[b]{0.24\textwidth}
         \centering
         \includegraphics[width=\textwidth]{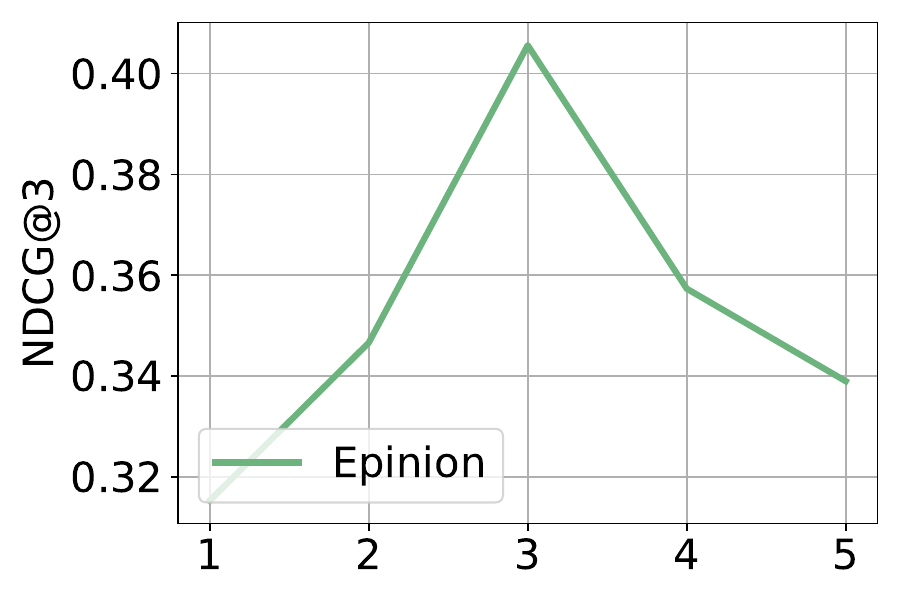}
         \caption{Number of layers $K$}
         \label{Epinion layer}
     \end{subfigure}
     \hfill
     \begin{subfigure}[b]{0.24\textwidth}
         \centering
         \includegraphics[width=\textwidth]{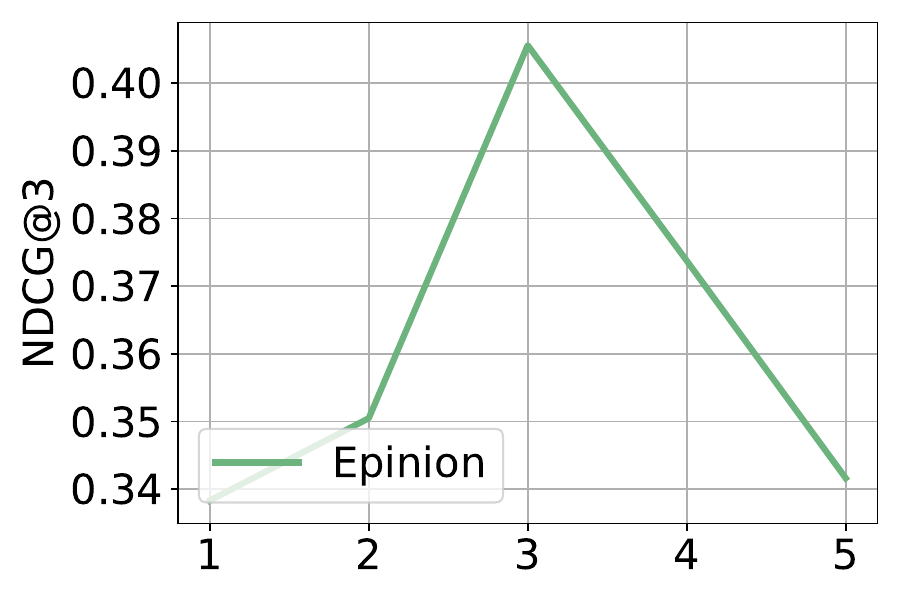}
         \caption{Number of heads $H$}
         \label{Epinion heads}
     \end{subfigure}
          \hfill
     \begin{subfigure}[b]{0.24\textwidth}
         \centering
         \includegraphics[width=\textwidth]{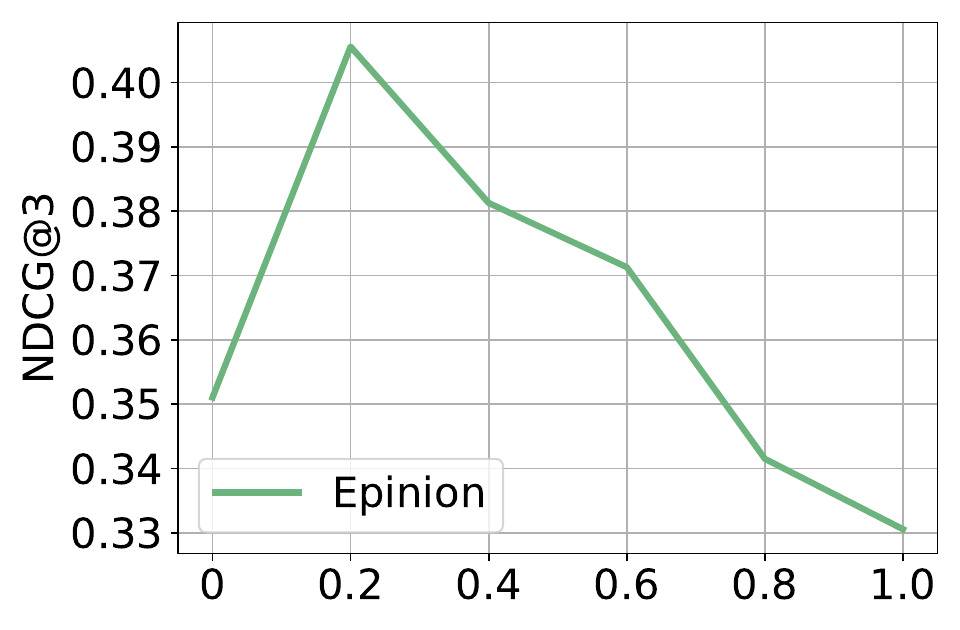}
         \caption{Regularization $\lambda_1$}
         \label{Epinion reg}
     \end{subfigure}

        \caption{The influence of model hyper-parameters $lr$, $K$, $L$, and $\lambda_1$.}
        \label{Parameters}
\end{figure*}

\subsection{RQ1: Performance Comparison}\label{sec:rq1}

Overall comparison results are shown in Table~\ref{tab:performance}. The best results are in boldface, and the second-best results are underlined. The improvement is calculated by subtracting the best performance value
of the baselines from that of MVPRec and then using the difference to
divide the former. We summarize the following key observations:
\begin{itemize}[leftmargin=*]
    \item Our proposed MVPRec method performs best and outperforms all the baseline methods in the three datasets. We hypothesize these large stable gains result from the capability of MVPRec to address the issue of role differentiation and heterogeneous information fusion.
    \item The simple MF method performs better than other baselines in Ciao and Epinion datasets, where per user has more social relations and more frequent group-buying behaviors. Compared with MF, GNN-based baselines applied on such heterogeneous graphs with denser edges may suffer from the over-smoothing problem, which leads to indistinguishable user representations. However, as a GNN-based model, MVPRec efficiently differentiates participant users from initiator users and thus shows even more obvious superiority over MF and other baselines in these two denser datasets.
    \item Although J2PRec is also a method designed for the GB problem, it neglects the distinction between the views of initiators and participants on items. Therefore, in more strict evaluation metrics, such as Recall@$K$ and NDCG@$K$ with $K\leq3$, it performs even worse than simpler methods like MF. In contrast, by distinguishing participants' views from initiators' views, MVPRec outperforms J2PRec by up to $101.44\%$ in all datasets. 
\end{itemize}

Since there’s no significant difference in terms of scalability and efficiency compared to the baseline graph convolution method, we do not include those experiments.

\begin{figure}
      \begin{center}
        \includegraphics[width=.24\textwidth]{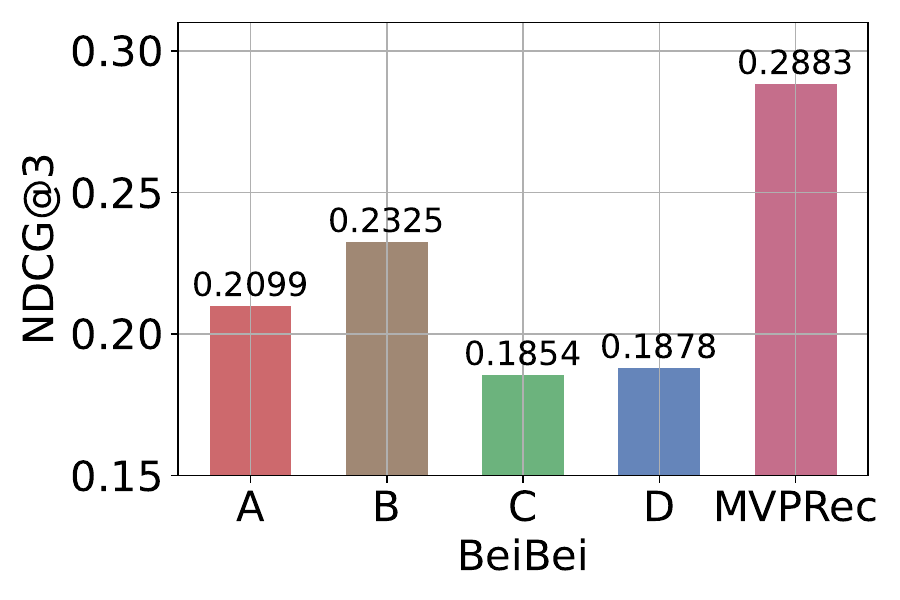}
        \includegraphics[width=.24\textwidth]{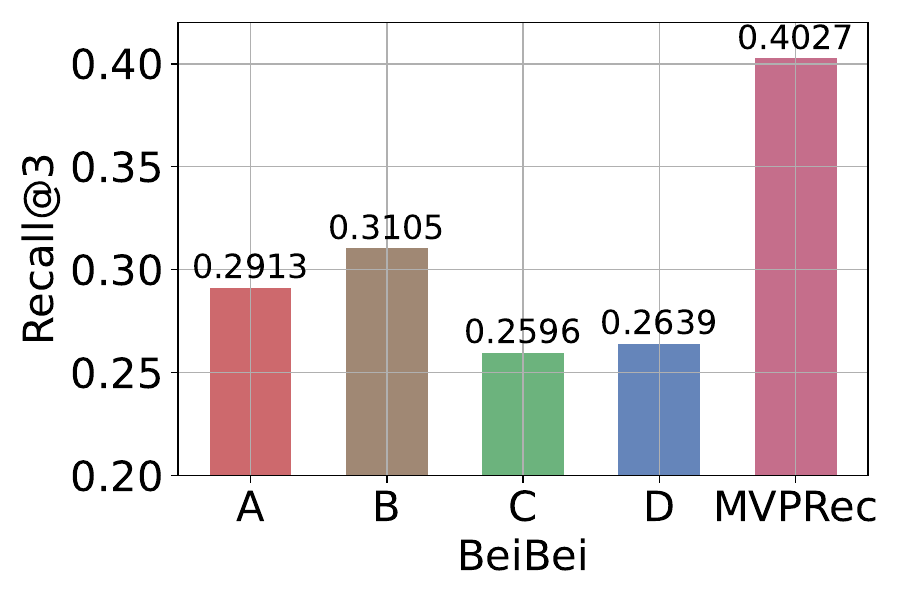}
        \includegraphics[width=.24\textwidth]{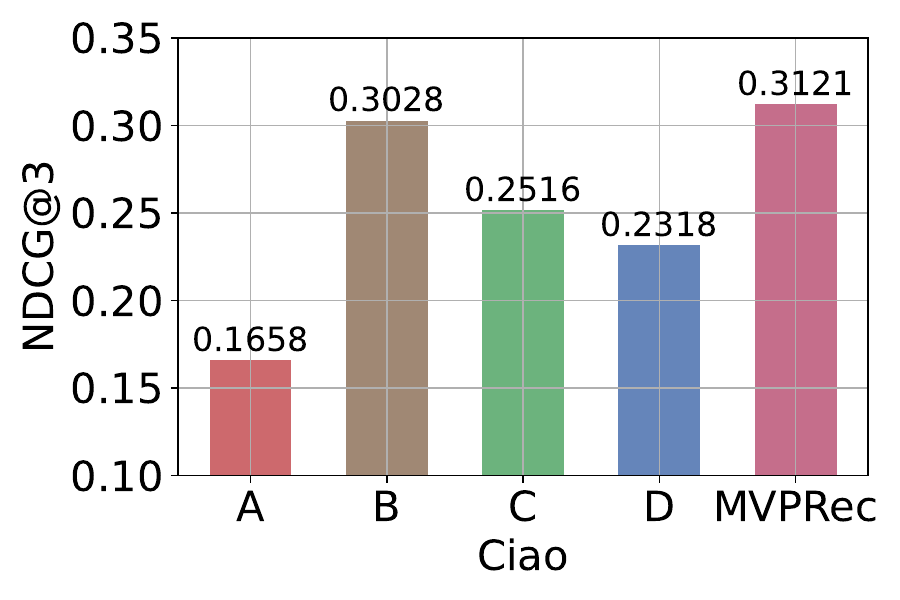}
        \includegraphics[width=.24\textwidth]{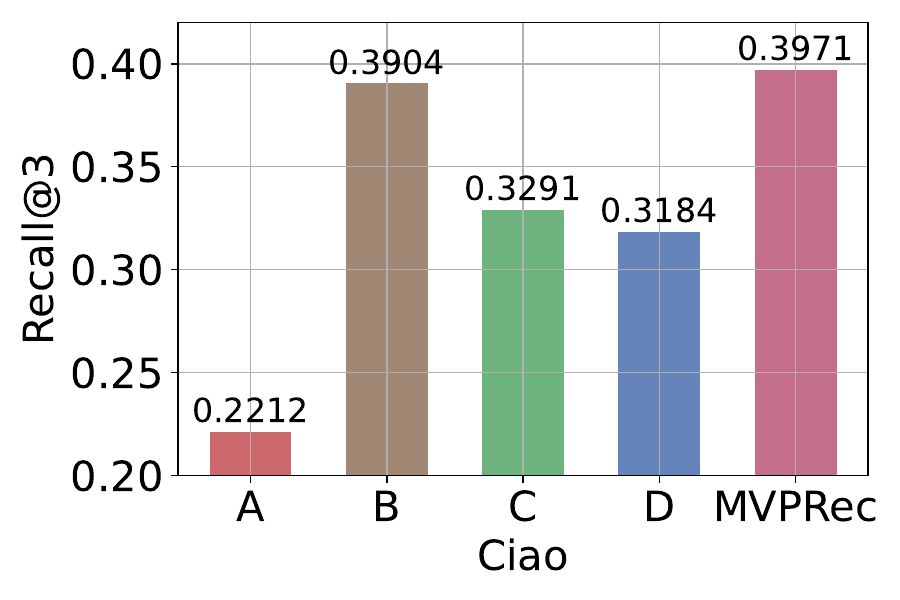}
        \includegraphics[width=.24\textwidth]{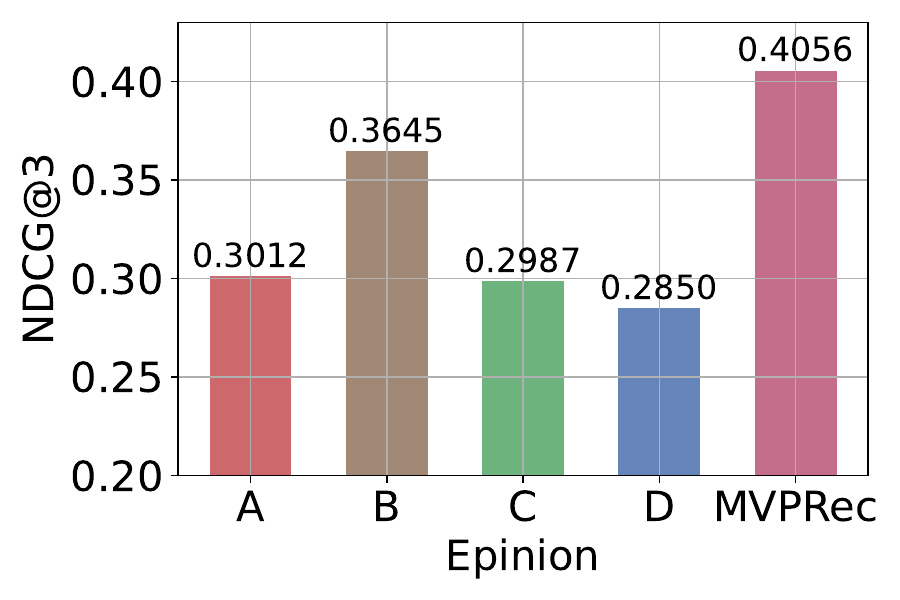}
        \includegraphics[width=.24\textwidth]{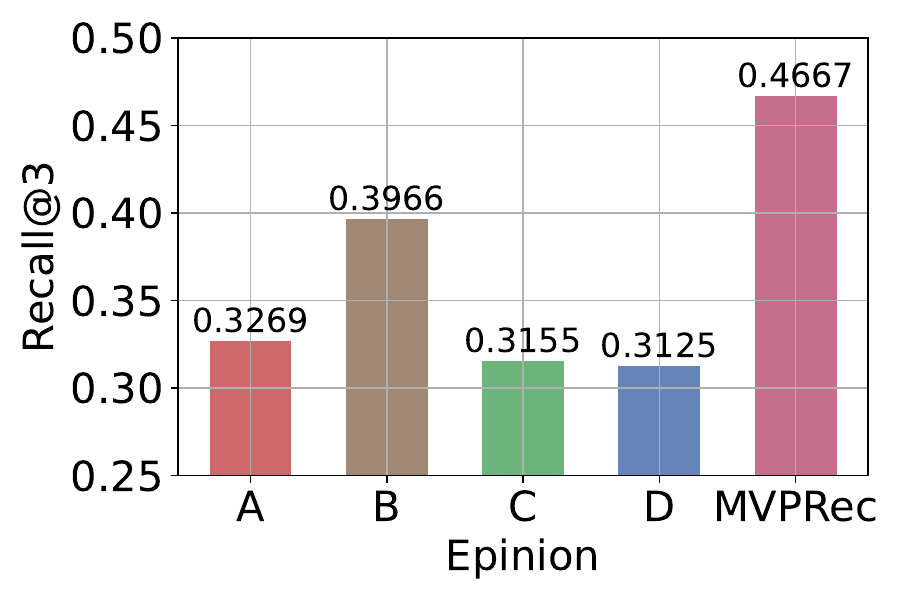}
      \end{center}
        \caption{Ablation study of \modelname.}
        \label{Ablation study}
\end{figure}

\subsection{RQ2: Ablation Study}
In this section, we perform an ablation study to evaluate the impact of each module on the overall framework on BeiBei and Ciao datasets.
We consider 4 variants: 1) Variant A replaces multi-view learning by unifying the initiator role and participant role, 2) Variant B abandons consistency loss on aligning item embeddings from two views, 3) Variant C is built by removing the multi-head attention mechanism, and 4) Variant D removes query strategy in Eq.\ref{query} and matches the participants directly with item embedding $\textbf{e}_i$. 
According to the comparison shown in Fig~\ref{Ablation study}, we have the following observations:
\begin{itemize}[leftmargin=*]
    \item  After removing the multi-view learning module, there are significant performance drops on both metrics, especially 46.88\% of NDCG@3 and 44.27\% of Recall@3 on Ciao dataset. 
    It greatly signifies the importance of modeling distinct views within the group buying behavior.
    \item The multi-head attention mechanism and query strategy play pivotal roles in determining the performance of \modelname on the BeiBei dataset. Their influences are significant, as evidenced by an average decrease of 35.6\% in both NDCG@3 and Recall@3 when these components are altered.
    \item The introduction of the consistency loss mechanism has the potential to further enhance the model's performance. This enhancement is evident, with a notable increase of 24\% in NDCG@3 and a substantial 29.69\% increment in Recall@3 on the BeiBei dataset.
\end{itemize}

\subsection{RQ3: Hyper-parameter Analysis}

In this section, we analyze the effect of four essential hyperparameters in \modelname by NDCG@3, including 1) learning rate, 2) number of layers $K$ in LGCN, 3) number of heads $H$ in multi-head attention mechanism, and 4) the coefficient on consistency loss $\lambda_1$. Experiment results are shown in Fig.~\ref{Parameters}.

The learning rate, a ubiquitous hyperparameter in many deep learning models, governs the rate at which an algorithm adapts its parameters. Based on the observations from Figure~\ref{BeiBei lr}, Figure~\ref{Ciao lr}, and Figure~\ref{Epinion lr}, it's apparent that the optimal learning rate values differ between datasets, and this parameter significantly impacts overall performance.
The hyperparameter $K$ is introduced in LGCN module and is an influential hyperparameter in the GNN-based recommender systems.
Stacking more layers, as we increase $K$, leads to capturing higher connectivities in the graph.
The performances on all datasets achieve the best results when $K = 3$.
Like the learning rate, the number of heads $H$ varies depending on the dataset in question.
The coefficient $\lambda_1$ regulates the level of consistency between item information in two distinct views. As depicted in Figure \ref{BeiBei reg}, Figure \ref{Ciao reg}, and Figure \ref{Epinion reg}, the model's performance generally shows an uptrend with increasing values of $\lambda_1$, but performance declines as this coefficient becomes excessively large.

\section{Related Works}
In this section, we provide a comprehensive review of various domains that are most relevant to GB, including collaborative filtering recommenders, social recommendation, group recommendation, and group-buying recommendation.
\subsection{Collaborative Filtering Recommenders}
With the assumption that users with similar interests are likely to have similar preferences toward items, collaborative filtering methods~\cite{CF} filter the interaction information between users and items and finally give the recommendation to the targeted users. The hybrid recommendation technique fuses various recommendation methods to make the best of both worlds. The model-based CF methods have been widely used in different recommendation scenarios. Matrix factorization (MF)~\cite{MF}, which projects users and items into latent vectors, models the historical interactions as the inner product of their projected vectors. Unlike MF, neural collaborative filtering (NCF)~\cite{NCF} models non-linearity via neural networks. 
LightGCN~\cite{LghtGCN} is a graph-based recommendation model focusing on simplicity and scalability. It just keeps the neighborhood aggregation part for collaborative filtering.
SGL~\cite{SGL} constructs contrastive views from the user-item bipartite graph by randomly dropping nodes and edges and maximizes the agreement between embeddings from contrasting perspectives to learn robust user and item embeddings. DirectAU~\cite{wang2022towards} directly aligns connected user-item pairs with a uniformity regularization. GraphAU~\cite{yang2023graph} further considers high-order collaborative filtering signal with alignment and uniformity loss, which is currently state-of-the-art method.

Collaborative filtering recommenders focus on learning user behavior patterns hidden behind observed user-item historical interactions, which is the commonly used data in current widely-applied recommender systems. This paper studies an emerging business mode (Group Buying) that greatly impacts the current E-commerce landscape. It is becoming prevalent with the E-commerce platform Temu and Pinduoduo. Different from collaborative filtering recommenders, our proposed MVPRec focuses on the new participant recommendation task underneath the group buying business model.

\subsection{Social Recommendation}
With the paths among users and edges between users and items, a random walk model TrustWalker~\cite{trustwalker} could provide confident recommendations with a collaborative filtering method. 
Based on MF, SocialMF incorporates social influence into learning latent vectors. Beexternals, many GNNs-based social recommendation models have achieved state-of-the-art performance. 
GraphRec~\cite{GraphRec} is the first work to apply GNNS to the social recommendation. It models the user preferences by combining the first-order aggregations from social and item neighbors via the attention mechanism. 
DiffNet~\cite{DiffNet} designs a layer-wise influence propagation layer to model user's latent vector by averaging the information of his/her friends. However, it neglects that collaborative interests in user-item interactions also play an essential role in learning user preferences. 
DiffNet++~\cite{DiffNet++}, an extension of DiffNet, is proposed to jointly model the high-order interests in both social networks and the user-item bipartite graph. 
MHCN~\cite{MHCN} combines hypergraph modeling and graph neural network to model high-order social relations with hyperedges. 
ConsisRec~\cite{yang2021consisrec} considers the inconsistency problem in social recommendation, and in each step it samples consistent social neighbors for aggregation. 
FeSoG~\cite{liu2022federated} studies the social recommendation task under the federated learning setting.
The disentangled social recommendation model DSR~\cite{DSR} first disentangles the user embeddings in the social networks into multiple facets and then updates final user embeddings via a facet-level attention mechanism. 
Considering user' preference is dynamic and influenced by friends, Song et al.~\cite{DGRec} design a dynamic attention module to capture user's current interests by dynamically inferring the friends with different influences.

Contrasting with the concept of social recommendation, GB entails a collective purchasing behavior involving multiple consumers. 
The key objective of participant recommendation is to identify potential friends who are likely to engage in a joint purchase of a targeted item alongside the initiator. 

\subsection{Group Recommendation}
Group recommendation aims to learn group representation by aggregating information from group members, and calculating the rating score from groups to items for recommendation. 
These models handle a group of users as a whole and ignore the distinctions of users in the same group.
AGREE~\cite{AGREE} adapts group representation by assigning attention weights to group members. MoSAN~\cite{MoSAN} models the user's preference with respect to all other members in the same group by sub-attention module. GroupIM~\cite{GroupIM} maximizes the mutual information between the user representation and their group representations which are aggregated from its members' preferences via the attention mechanism. Further, Zhang et al.~\cite{double-scale} design a double-scale node dropout strategy to generate self-supervision signals to alleviate the data sparsity issue and capture both the intra- and inter-group interactions among users. 
Suggesting groups to potential users is a highly relevant task that has attracted considerable attention from both the academic and industrial sectors. 
CFAG~\cite{CFAG} devises tripartite graph convolution layers to aggregate information from different types of neighborhoods (users, items, and groups) and proposes a propagation augmentation (PA) layer to mitigate data sparsity.
GTGS~\cite{yang2023group} incorporates a THC layer to transfer item preferences from members to groups, ensuring user interests contribute to group identification, and employs Cross-view Self-Supervised Learning (CSSL) to maintain consistency between item and group preferences for each user.

The data format employed in group recommendation, often encompassing social relations, bears resemblance to that used in GB.
However, a distinctive aspect of GB is the dual role framework it offers users: they can function either as initiators or participants. 
This is a notable departure from the conventional single-role paradigm prevalent in social networks and groups.

\subsection{Group Buying Recommendation}
GBGCN~\cite{GBGCN} designs in-view (initiator view and participant view) propagation and cross-view (user in initiator view, user in participant view, item in initiator view, and item in participant view) propagation to let embedding capture the high-order information and the preferences of different roles, and finally predicts the probability of a user launching a successful group buying with the item in data by combining the preferences of initiator and participants over that item. SHGCN~\cite{SHGCN} aims to enhance personalized recommendations by building hypergraph for each group-buying data with regarding the item as the hyperedge to capture the inhomogeneous social influence and learn the user preferences in a fine-grained manner. 
Different from friend recommendation which recommends similar users to the target user, participant recommendation needs to take the initiator and item into consideration since they both have influences on the decision of potential participants. J2PRec~\cite{J2PRec} proposes a joint learning framework to recommend both items and participants to maximize the GB likelihood. It updates user and item embedding by aggregating from relational group-buying graph under a probabilistic framework on both recommendation tasks. 

In contrast to prior research in the domain of GB, our model, referred to as \modelname, places particular emphasis on the core task of recommending prospective participants from one's social network, a fundamental function within GB platforms.

\section{Conclusion}
With the prevalence of group buying in many e-commerce platforms, the participant recommendation task is still under exploration.
This study introduces a novel framework, \modelname, designed to enhance participant recommendations in the context of group buying.
We introduce the multi-view learning module to differentiate the roles of users (Initiator/Participant) with the GB.
We construct a multi-view user representation through graph encoders by fusing these two views with the social graph. 
Subsequently, MVPRec seamlessly integrates these GB and social representations, leveraging an attention module to synthesize a comprehensive user representation. 
Furthermore, the model harnesses the power of a multi-head attention mechanism to learn matching scores by considering the social connections between Initiators and their friends. 
This multifaceted approach enhances the efficacy of participant recommendations in group buying scenarios.

\section{Acknowledge}
This work is supported in part by NSF under grant III-2106758. 

\bibliographystyle{IEEEtran}
\balance
\bibliography{ref_new}

\end{document}